\documentclass[fleqn,10pt]{wlscirep}
\usepackage[utf8]{inputenc}
\usepackage[T1]{fontenc}
\usepackage{subfigure}
\usepackage{tikz}
\usetikzlibrary{shapes.geometric}
\usepackage{braket}

\title{Tensor Network Machine Learning for Wildfire Susceptibility Mapping: from Grokking Dynamics to Quantum Mixedness of Class Representations}

\author[1,2,$\dagger$]{Domenico Pomarico}
\author[1,$\dagger$]{Alessandra Costantino}
\author[1,$\dagger$]{Gabriel Ramirez Sanchez}
\author[2,3]{Loredana Bellantuono}
\author[4]{Davide D'Alò}
\author[4]{Mario Elia}
\author[1,2]{Alessandro Fania}
\author[1,2]{Francesco Giordano}
\author[1]{Niloofar Kheirkhahan}
\author[4]{Raffaele Lafortezza}
\author[1,2,*]{Ester Pantaleo}
\author[2,4]{Sabina Tangaro}
\author[1,2]{Roberto Bellotti}
\author[1,2,+]{Alfonso Monaco}
\author[2,5,+]{Nicola Amoroso}
\affil[1]{Dipartimento Interuniversitario di Fisica, Universit\`a degli Studi di Bari Aldo Moro, Bari, I-70126, Italy}
\affil[2]{Istituto Nazionale di Fisica Nucleare, Sezione di Bari, Bari, I-70126, Italy}
\affil[3]{Dipartimento di Biomedicina Traslazionale e Neuroscienze (DiBraiN), Universit\`a degli Studi di Bari Aldo Moro, Bari, I-70124, Italy}
\affil[4]{Dipartimento Di Scienze Del Suolo, Della Pianta e Degli Alimenti (DISSPA), Universit\`a degli Studi di Bari Aldo Moro, Bari, I-70125, Italy}
\affil[5]{Dipartimento di Farmacia-Scienze del Farmaco, Universit\`a degli Studi di Bari Aldo Moro, Bari, I-70125, Italy}

\affil[*]{ester.pantaleo@uniba.it}

\affil[$\dagger$]{these authors contributed equally to this work}

\affil[+]{these authors contributed equally to this work}

\begin{abstract}
A quantum-inspired tensor network framework for wildfire susceptibility classification in the Gargano region is introduced, leveraging AlphaEarth embeddings and Matrix Product State models. The approach combines scalable geospatial representations with an interpretable quantum mask, enabling both binary and multiclass classification of wildfire susceptibility. Beyond predictive performance, the study reveals a pronounced grokking transition in the binary case and provides a detailed analysis of inter-class confusion in the multiclass setting. By introducing level-resolved mixedness diagnostics based on reduced density matrices, we show that the MPS classifier naturally encodes a hierarchy of class distinguishability, with non-adjacent categories becoming more separable than neighboring ones. These results demonstrate that tensor network models not only achieve competitive classification accuracy but also offer a physically grounded framework to quantify and interpret class separability in complex environmental datasets.
\end{abstract}
\begin{document}

\pagecolor{white}

\flushbottom
\maketitle
\thispagestyle{empty}

%\noindent \textcolor{blue}{Please note: Abbreviations should be introduced at the first mention in the main text – no abbreviations lists. Suggested structure of main text (not enforced) is provided below.}

\section*{Introduction}

The continuous expansion of satellite-based Earth Observation (EO) systems has transformed EO into a data-intensive discipline, where large-scale and heterogeneous datasets are routinely generated. Modern missions, equipped with multispectral and hyperspectral sensors and high temporal resolution, produce massive streams of geospatial information that require advanced methods for efficient processing and analysis. Artificial intelligence, and in particular machine learning (ML) and deep learning, have become essential tools to extract knowledge from such data. However, classical approaches face increasing pressure in terms of computational cost, scalability, and the ability to capture complex spatial-spectral relationships inherent in EO data \cite{Sebastianelli2025QMLReview}.

In this context, quantum computing (QC) has emerged as a promising paradigm that leverages quantum mechanical phenomena such as superposition and entanglement to process information in high-dimensional spaces. The intersection between QC and ML has led to the development of quantum machine learning (QML), which aims to enhance learning processes by exploiting quantum representations. Although QML shows potential improvements in optimization efficiency and representation learning, current devices are still constrained by noise, limited qubit availability, and short coherence times. As a result, most practical implementations rely on hybrid or quantum-inspired approaches \cite{Sebastianelli2023Hyperparameters, Sebastianelli2022QCNN}.

A crucial aspect in the application of advanced learning models to EO data is the construction of meaningful input representations. This issue is becoming increasingly relevant as EO research moves toward large-scale, cloud-native and multi-source geospatial analysis. The rapid growth of studies based on platforms such as Google Earth Engine, together with the recent emergence of GeoAI-driven workflows, satellite embeddings, multimodal data fusion and foundation-model-based approaches, highlights the expansion of this research direction. In this context, representation learning is becoming central for transforming heterogeneous EO observations into compact and informative features suitable for mapping and prediction tasks. For instance, AlphaEarth Foundations \cite{Brown2025AlphaEarth} proposes a global embedding field that integrates heterogeneous environmental information into a compact representation, enabling accurate mapping even with sparse label data.

In this work, we leverage Google Earth Engine (GEE) as a scalable platform for satellite data processing and feature extraction. GEE enables the efficient integration of multi-source remote sensing datasets, such as multispectral imagery, vegetation indices, topographic variables, and climatic indicators, into a unified framework\cite{Khachoo2026GEE}. These data are preprocessed and transformed into structured feature maps, which we treat as an embedding of the environmental state. This embedding step condenses high-dimensional satellite observations into a compact representation that can be effectively processed by downstream machine learning models.

Building on this representation, we adopt quantum-inspired tensor network models as the core learning architecture. Tensor networks provide a powerful framework to encode high-dimensional correlations through low-rank decompositions, mimicking the entanglement structure of quantum systems while remaining fully implementable on classical hardware. In this sense, the GEE-based embedding can be interpreted as the classical input layer, while the tensor network acts as a structured model capable of capturing nonlinear interactions across multiple spectral dimensions.

The Mediterranean Basin is widely recognized as a climate-change hotspot, where rising temperatures, recurrent droughts, summer heat extremes, and changes in precipitation patterns are increasing environmental stress and altering fire-prone conditions\cite{UrdialesFlores2023,Lazoglou2024}. Mediterranean landscapes are also highly diverse, making the identification of high-vulnerabil areas a complex task that requires models capable of accounting for heterogeneous and interacting environmental drivers.
This complexity is evident in Southern Italy, where strong spatial variability in vegetation structure, land use, terrain morphology, human pressure, and seasonal fuel availability jointly shape wildfire-prone conditions\cite{cilli2022,wildfire2026}.
Accurate mapping of wildfire susceptibility in such conditions requires the integration of multi-source geospatial information and the modeling of complex environmental interactions, thereby motivating the adoption of advanced representation learning techniques, such as Matrix Product State (MPS) classifiers.

In this work, we focus on the application of these methods to the identification of areas of high wildfire susceptibility in one specific region of Southern Italy: the Gargano promontory. This area represents a particularly challenging Mediterranean case study, owing to the spatial heterogeneity of its landscape and the combined influence of environmental and human-related fire drivers.

The study is structured as follows. In the Materials section, we introduce the wildfire reference data through a four-class label map derived for the Gargano area and present the AlphaEarth Foundations embedding framework as a compact, multi-source representation of environmental variability. In the Results section, we first analyze the emergence of a grokking transition in the binary classification task (class 0 versus the aggregated classes 1, 2, and 3), highlighting the delayed yet abrupt gain in generalization observed in the MPS classifier. We then extend the analysis to the four-class setting, where we provide a detailed investigation of class-dependent performance and the role of confounding classes in limiting generalization. Finally, in the Discussion section, we compare the proposed approach with the current state of the art in wildfire susceptibility modeling, emphasizing both the advantages and limitations of quantum-inspired tensor network models in EO applications.

\begin{figure*}[t!]
\subfigure[]{\centering
\begin{tikzpicture}
	\node[inner sep=0pt] (russell) at (-0.45,-5) {\includegraphics[width=0.35\linewidth]{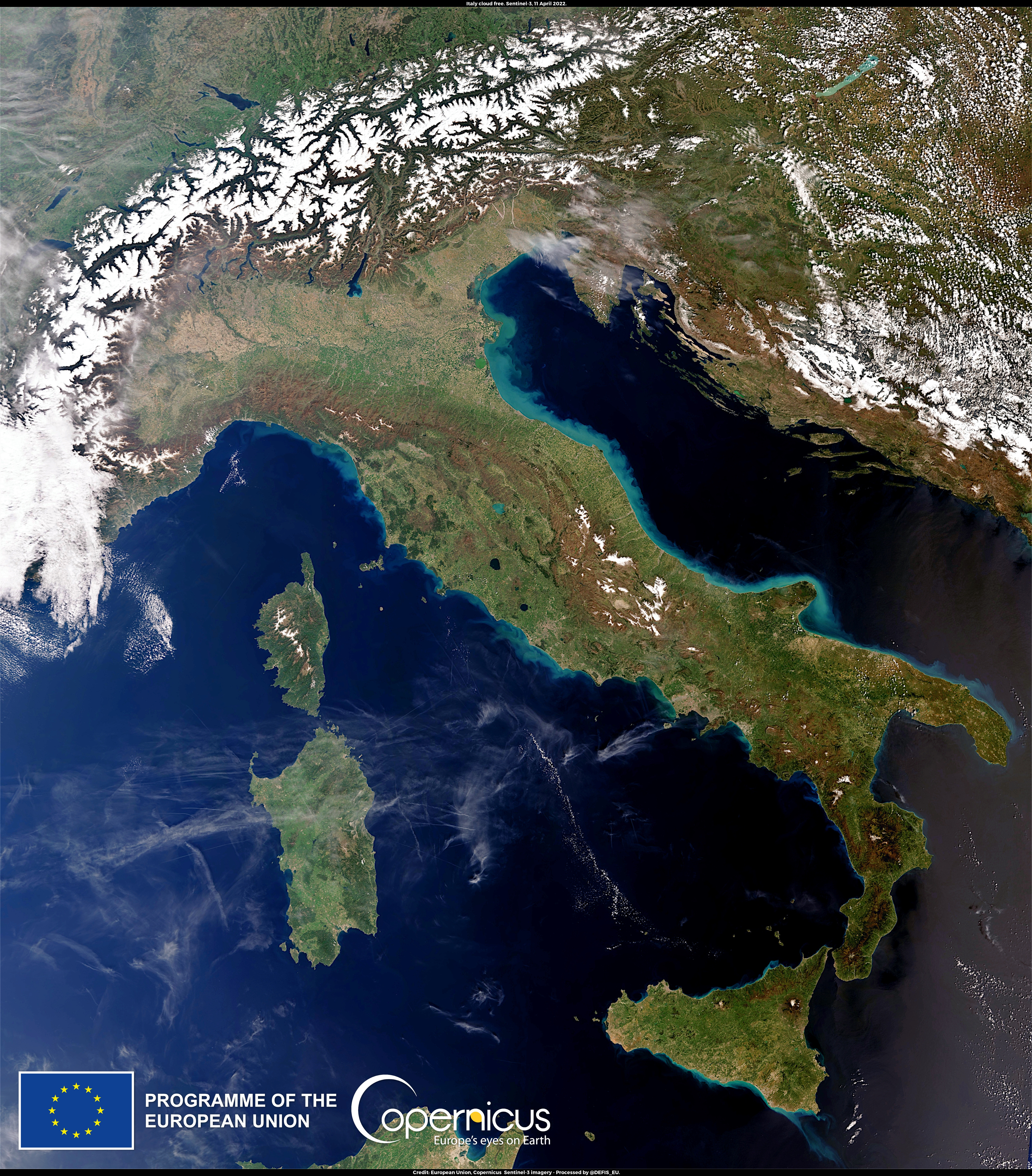}};
	\node[inner sep=0pt] (russell2) at (9,-5) {\includegraphics[clip, width = 0.59\linewidth{}]{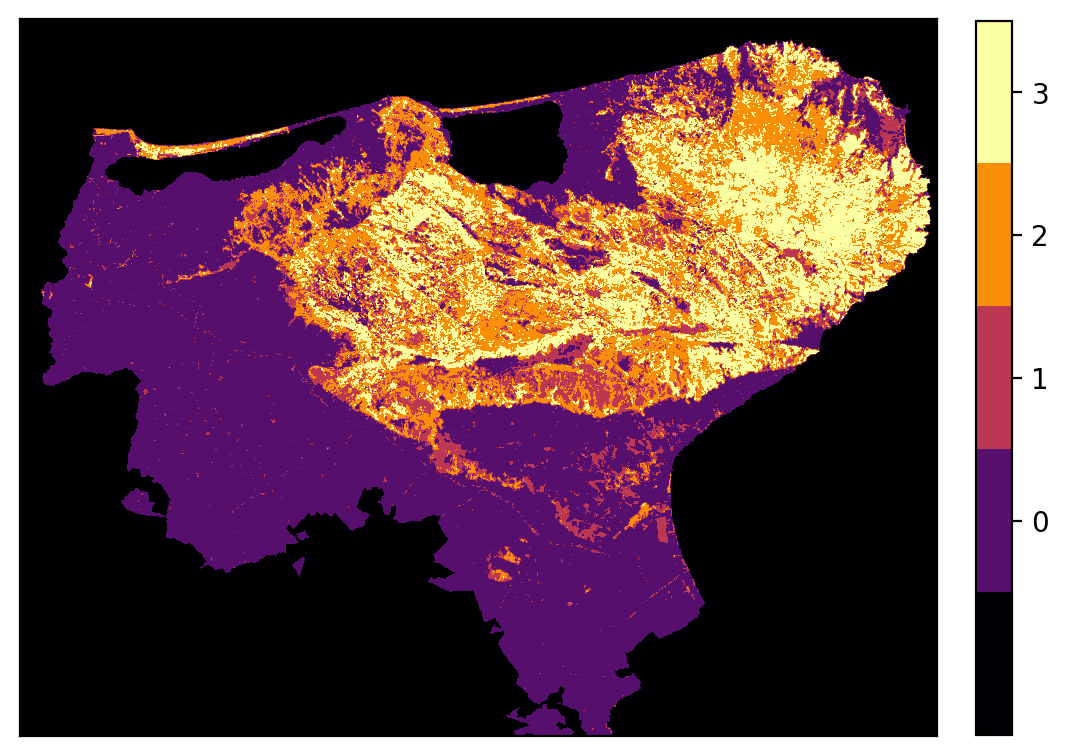}};
	\draw[-, gray, line width=0.3mm] (0.9,-4.94) -- (1.37,-4.94) -- (1.37,-5.28) -- (0.9,-5.28) -- (0.9,-4.94);
	\draw[-, gray, dashed, line width=0.3mm] (0.9,-5.28) -- (4,-8.45);
	\draw[-, gray, dashed, line width=0.3mm] (0.9,-4.94) -- (4,-1.5);
\end{tikzpicture}} \\
\subfigure[]{\centering
\begin{tikzpicture}
	\node[inner sep=0pt] (russell) at (0.85,-5) {\includegraphics[width=0.35\linewidth]{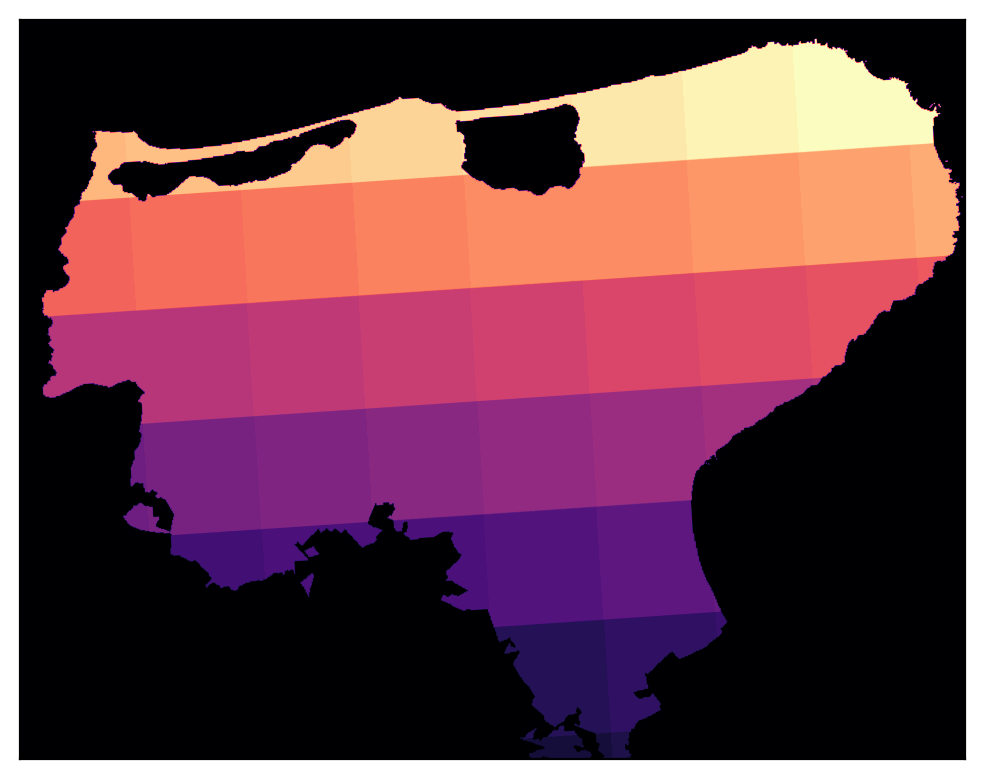}};
	\draw[->,thick] (-2.4,-3.7) to [bend right=270] (-2.4,-1);
    \node[] (t8) at (-4.15,-2.35) {\rotatebox{90}{embedding pixels}};
    \node[] (t8) at (-3.65,-2.35) {\rotatebox{90}{sampling from super-pixels}};
    %\draw[->, line width=0.3mm] (-1,-1) -- (-0.5,-1);
    \node[inner sep=0pt] (russell2) at (-1.75,-1) {\includegraphics[clip, width = 0.05\linewidth{}]{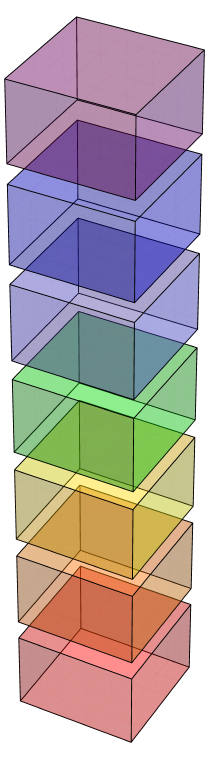}};
    \draw[<->,thick] (-0.7,-0.3) to [bend right=90] (-0.7,-1.7);
    \draw[rounded corners,yellow!40,fill] (-0.4, 0.25) rectangle (2.2, -0.85);
    \draw[rounded corners,orange!40,fill] (-0.4, -1.15) rectangle (2.2, -2.25);
    \node[] (t8) at (0.9,-1.7) {$4$ classes};
    %\node[] (t9) at (1.1,-1.95) {$0$ vs $1$ vs $2$ vs $3$};
    \node[] (t8) at (0.9,-0.05) {$2$ classes};
    \node[] (t9) at (0.9,-0.55) {$0$ vs $\{1,2,3\}$};
    \draw[->,thick] (2.4,-0.4) -- (3,-0.4);
    \draw[->,thick] (2.4,-1.6) -- (3,-1.6);
    \draw[rounded corners,red!50,fill] (3.2, -0.25) rectangle (6.1, -1.75);
    \node[] (t8) at (4.65,-0.45) {boruta features};
    \node[] (t8) at (4.65,-0.95) {selection for};
    \node[] (t8) at (4.65,-1.45) {wildfire classes};
    %\draw[rounded corners,brown!30,fill] (10, 0.3) rectangle (13, -0.7);
    %\draw[<-, line width=0.3mm] (12,-1) -- (12,-1.5);
    %\node[] (t8) at (11.5,0.1) {coarse grained};
    %\node[] (t8) at (11.5,-0.4) {resolution $6\times 6$};
    \draw[rounded corners=3pt,line width=0.3mm,->]
    (6.35,-1) -- (8.85,-1) |- (8.85,-1.4);
    %\draw[rounded corners=3pt,line width=0.3mm,->]
    %(9.65,0) -- (7.2,0) |- (7.2,-0.4);
    \draw[rounded corners,purple!40,fill] (6.5, -1.6) rectangle (11.2, -2.4);
    \node[] (t8) at (8.85,-2) {MPS quantum mask training};
    %\draw[->, line width=0.3mm] (9.35,-2.6) -- (9.35,-3.15);
    \draw[<->,thick] (9.85,-3.15) to [bend right=90] (7.85,-3.15);
    %\node[inner sep=0pt] (russell) at (6.5,-3.5) {\includegraphics[clip, width = 0.3\linewidth{}]{images/entanglement_transition.png}};
    \draw[rounded corners,blue!40,fill] (4.7, -3.4) rectangle (8.7, -4.3);
    \node[] (t8) at (6.7,-3.6) {grokking transition};
    \node[] (t8) at (6.7,-4.1) {performance evaluation};
    \node[inner sep=0pt] (russell3) at (6.3,-6.2) {\includegraphics[clip, width = 0.3\linewidth{}]{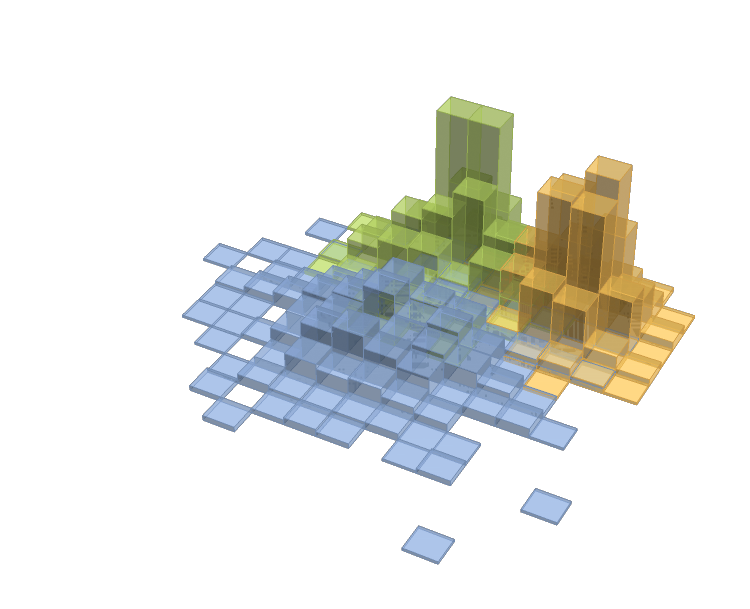}};
    \draw[rounded corners,green!40,fill] (9, -3.4) rectangle (13, -4.3);
    \node[] (t8) at (11,-3.6) {confounding classes};
    \node[] (t8) at (11,-4.1) {quantum masks mixedness};
    \node[inner sep=0pt] (russell3) at (11,-5.8) {\includegraphics[clip, width = 0.15\linewidth{}]{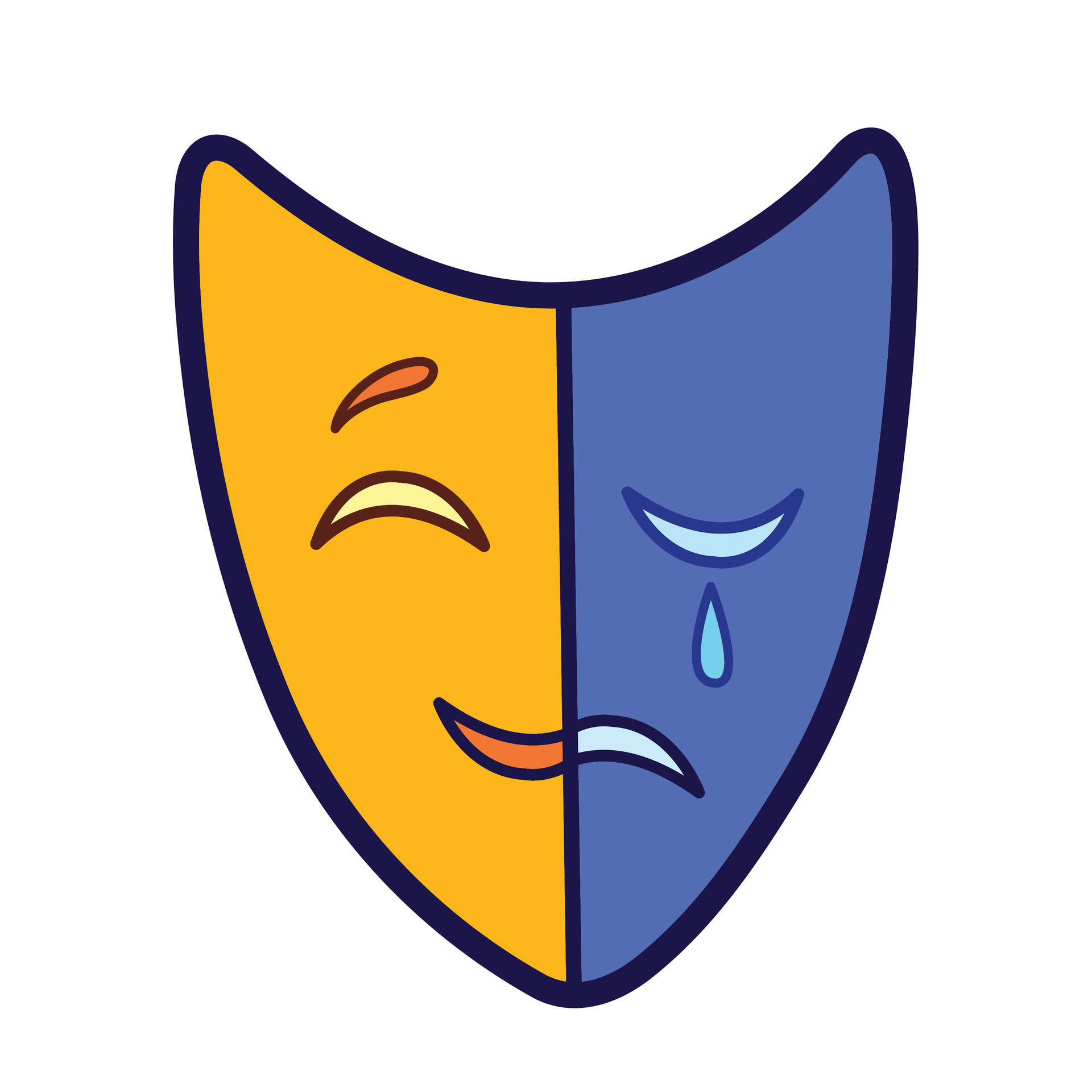}};
\end{tikzpicture}}
\caption{Study area and workflow for wildfire classification in the Gargano region. (a) Location of the study area within southern Italy and corresponding four-class wildfire susceptibility map at pixel resolution. (b) Schematic pipeline illustrating the group-fold sampling of embedding features from graded color super-pixels, binary (0 vs {1,2,3}) and multiclass formulations, feature selection, MPS-based quantum mask training leading to grokking-driven performance evaluation, and detection of mixed quantum masks with underlying pattern recognition or characterization of not separable confounding labels. \label{fig:gargano}}
\end{figure*}
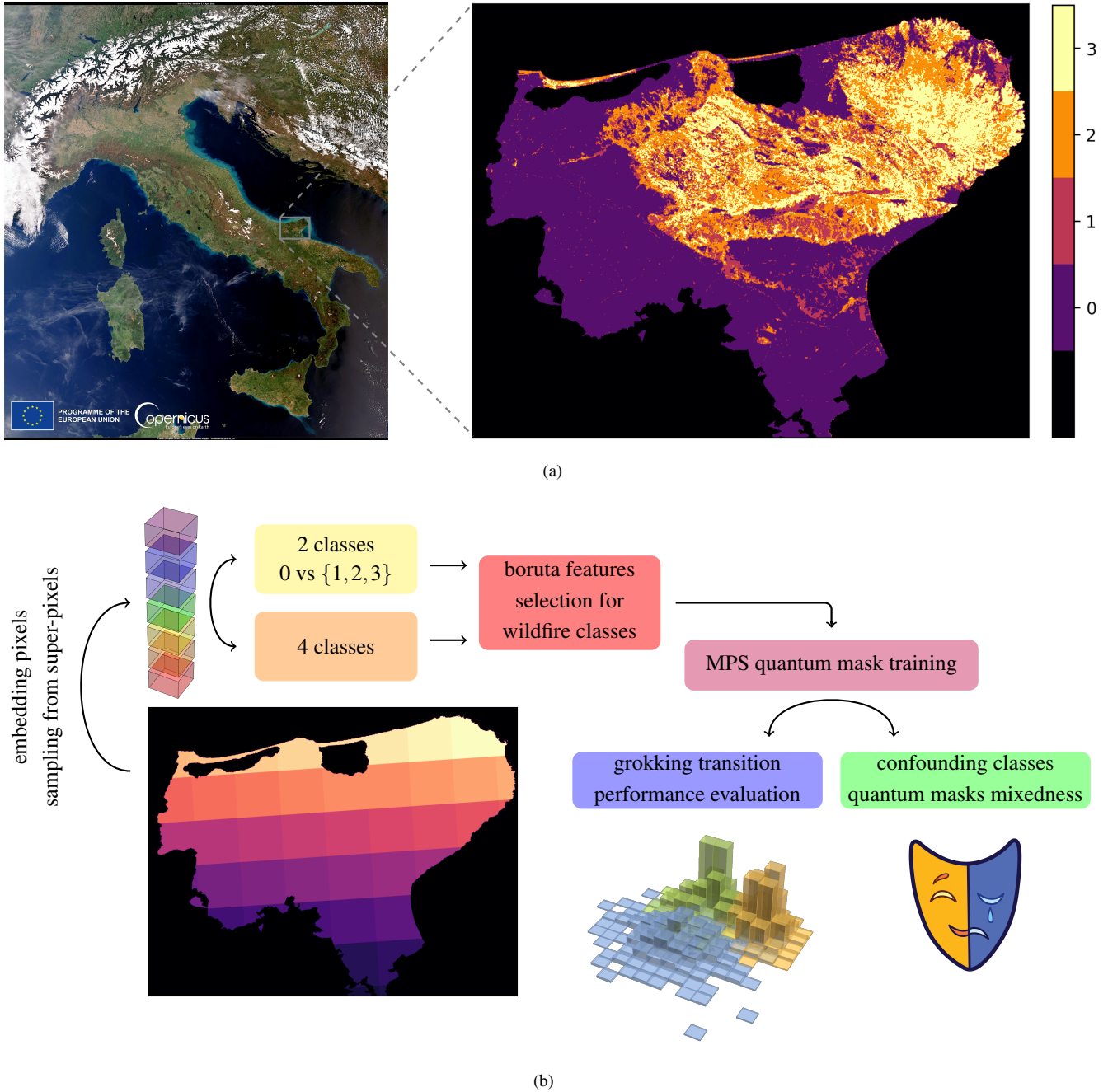

\section*{Materials}

\subsection*{Study area and wildfire reference data}

The Gargano Region, located in northern Puglia, southern Italy, was selected as the study area and is shown in Figure \ref{fig:gargano}(a). This area is characterized by a heterogeneous landscape comprising forested uplands, agricultural zones, and coastal urban settlements. The region has historically been affected by recurrent wildfire events, driven by the combined influence of Mediterranean climatic conditions, fuel availability, and anthropogenic pressures.
The wildfire target layer was generated by combining two complementary components: a fuel hazard layer and an ignition danger layer. The hazard component was derived from predicted 1-hour fine dead fuel load, estimated from multi-source remote-sensing and ancillary data \cite{DEste2021Fuel}. The danger component was based on environmental, land-cover, and anthropogenic predictors, together with historical wildfire occurrence data, in line with previous wildfire probability modeling approaches for Mediterranean landscapes \cite{Elia2020Wildfire}. The final wildfire susceptibility map, obtained by combining (multiplying) the predicted fuel load and danger maps, thus integrating information on fuel availability and ignition probability\cite{wildfire2026}, after reclassification into a four-class label map to capture increasing levels of wildfire vulnerability. 
The label maps were downsampled to a spatial resolution of $10 \, m \times 10 \, m$ grid cells, where each cell represents a single observation unit for model training.

In this framework, the four classes can be interpreted as:
\begin{itemize}
\item Class 0, no/negligible susceptibility: areas with minimal fuel load and/or unfavorable ignition conditions (e.g., urban cores, water bodies).
\item Class 1, low susceptibility: areas with limited fuel continuity or moderate environmental constraints on fire spread.
\item Class 2, moderate susceptibility: areas where both vegetation structure and climatic conditions allow for recurrent fire occurrence.
\item Class 3, high susceptibility: zones characterized by dense fuel availability and favorable ignition conditions, often associated with forested and shrub-dominated landscapes.
\end{itemize}

This discretization allows capturing gradients in wildfire processes while preserving interpretability within machine learning frameworks. The use of a regular grid ensures scalability and compatibility with satellite-derived predictors and geospatial embeddings. In the broader modeling workflow, wildfire occurrence in the Gargano region was strongly associated with fuel availability, land cover composition (e.g., forest, grassland), and the Fire Climate Index, which emerged as the dominant explanatory variables in the study area.

\subsection*{AlphaEarth embedding field representation}

This study leverages AlphaEarth Foundation's embeddings as the main geospatial representation for wildfire susceptibility modelling. An embedding is a learned numerical representation that transforms complex input data into a vector of features in a latent space, where locations with similar environmental characteristics are expected to have similar vector representations \cite{Bengio2013RepresentationLearning}. AlphaEarth Foundations extends this idea to EO by producing analysis-ready embedding fields over the Earth’s surface. Each $10 \, m \times 10 \, m$ pixel is represented by a 64-dimensional vector that summarizes information derived from multiple satellite and environmental data sources over an annual time window \cite{Brown2025AlphaEarth}. Therefore, instead of explicitly constructing separate predictor layers for land cover, vegetation dynamics, surface conditions, and other environmental descriptors, the model can directly use this unified latent representation as input.
This approach is particularly relevant for wildfire susceptibility modelling because fire occurrence is controlled by the interaction of several factors, including fuel availability, vegetation dynamics, climatic seasonality, landscape configuration, and anthropogenic pressure. Using AlphaEarth embeddings allows these heterogeneous influences to be represented in a compact and consistent feature space, while reducing the burden associated with data collection, preprocessing, temporal compositing, spatial resampling, and cross-dataset harmonization. The resulting representation provides a scalable basis for susceptibility mapping in the Gargano Urban Region and may support transfer to other Mediterranean fire-prone landscapes characterized by strong environmental heterogeneity.
The four-class wildfire label map for the Gargano area was used as the target variable, while the AlphaEarth embedding dimensions were used as input features for the downstream machine-learning models. To ensure temporal consistency with the reference wildfire layers, which were generated from 2019 satellite imagery, ancillary datasets, and field inventory data, the corresponding annual AlphaEarth embeddings were downloaded and used as input features. Since the embedding vector contains 64 latent features for each pixel, Boruta feature selection was applied to identify the most informative dimensions and reduce feature redundancy before model training \cite{Boruta2010}. This workflow links an interpretable target representation, expressed through discrete wildfire susceptibility classes, with a compact data-driven description of environmental conditions provided by AlphaEarth. As a result, the approach supports wildfire hazard classification while reducing the need for manual predictor engineering and facilitating scalable application to larger areas or multi-hazard monitoring contexts.

\section*{Results}

\begin{figure*}[t!]
\includegraphics[width=\linewidth]{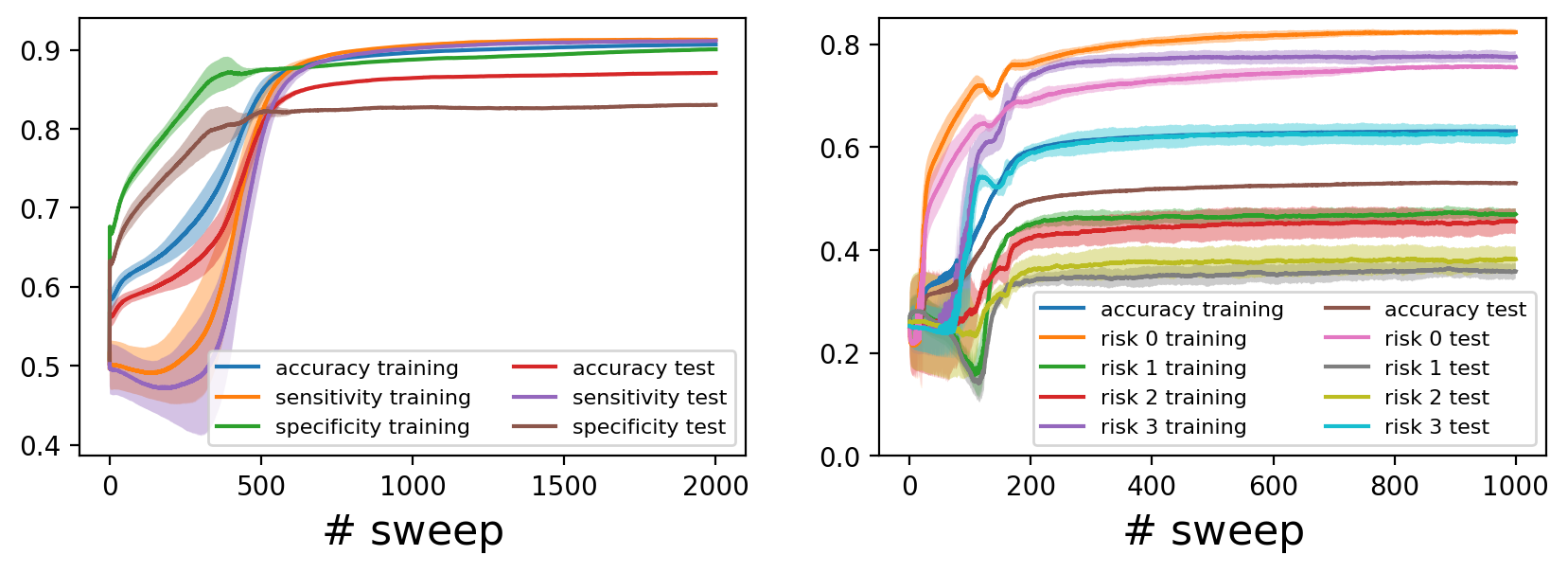}
\caption{Evolution of training and test performance of the MPS classifier as a function of training sweeps. The left panel is referred to binary classification (class 0 vs classes 1–3), the right panel to the four-class classification. Solid lines denote the mean values, while shaded regions represent the corresponding standard deviation computed over three independent runs. \label{fig:performance}}
\end{figure*}

The proposed workflow is summarized in Figure \ref{fig:gargano}(b), which illustrates the transition from AlphaEarth embeddings to Boruta-selected features and downstream wildfire susceptibility classification. To mitigate the effect of spatial autocorrelation, a well-known issue in EO classification tasks, we construct the training and test datasets by sampling $10^4$ embedding pixels from a set of super-pixels, each aggregating nearly $10^6$ original pixels, in a balanced way with respect to the associated classes. We partition super-pixels according to $70\%$ for training and $30\%$ for testing, respectively. This strategy ensures that sampled observations are spatially decorrelated, reducing the risk of information leakage between training and test sets and providing a more reliable assessment of generalization performance.

\begin{table}[b!]
\centering
\caption{Classification performance metrics reported as mean $\pm$ standard deviation.}
\label{tab:classification_metrics}
\begin{tabular}{lc}
\toprule
\textbf{Metric} & \textbf{Mean $\pm$ Std} \\
\midrule
\multicolumn{2}{c}{\textbf{Binary Classification}} \\
\midrule
Accuracy    & $87.07 \, \% \pm 0.04 \, \%$ \\
Sensitivity & $91.13 \, \% \pm 0.08 \, \%$ \\
Specificity & $83.02 \, \% \pm 0.09 \, \%$ \\
\midrule
\multicolumn{2}{c}{\textbf{Multiclass Classification}} \\
\midrule
Accuracy & $53.01 \, \% \pm 0.20 \, \%$ \\
Class 0  & $75.47 \, \% \pm 0.47 \, \%$ \\
Class 1  & $35.88 \, \% \pm 1.52 \, \%$ \\
Class 2  & $38.21 \, \% \pm 2.52 \, \%$ \\
Class 3  & $62.49 \, \% \pm 1.73 \, \%$ \\
\bottomrule
\end{tabular}
\end{table}

On this decorrelated dataset, we apply a Boruta feature selection procedure to identify the most relevant predictors among the available embedding features. The selection process yields a consistent subset of $30$ informative features, as reported in Methods, which are subsequently used for both the binary classification task (class 0 versus classes 1–3) and the full four-class classification problem. This unified feature set enables a direct comparison between the two formulations, allowing us to isolate the effects of model dynamics, such as the emergence of grokking, from confounding influences related to feature dimensionality or input variability.

The analysis was structured around two complementary classification formulations: a binary task, separating class 0 from classes 1--3, and the original four-class task. The binary formulation provides both a coarse assessment of wildfire susceptibility and a benchmark against the classical modeling approach reported in \cite{wildfire2026}, whereas the four-class formulation evaluates the ability to capture finer hazard gradients. Both tasks were trained using the same Boruta-selected set of 30 AlphaEarth embedding features, enabling a direct comparison of model behavior under identical input conditions.
The binary classification metrics reported in Table \ref{tab:classification_metrics} show a strong performance, with an accuracy of $87.07\% \pm 0.04\%$, high sensitivity $91.13\% \pm 0.08\%$, and slightly lower specificity $83.02\% \pm 0.09\%$. The four-class task was more challenging, reaching $53.01\% \pm 0.20\%$ accuracy, with better performance for classes 0 and 3 than for the intermediate classes 1 and 2.
Beyond the final performance scores, the training trajectories provide insight into how these results were achieved. 
The training dynamics shown in Figure \ref{fig:performance} reveal a clear manifestation of a grokking-like transition in the MPS classifier \cite{pomarico2025grokkingentanglementtransitiontensor,technologies13100438}, particularly evident in the binary formulation (class 0 versus classes 1–3). In the early stages of training, the model displays a characteristic gap between training and test performance: training accuracy increases steadily, while test accuracy and sensitivity remain significantly lower, indicating that the model is primarily capturing superficial correlations or class biases rather than learning a generalizable decision boundary. During this phase, specificity rises more rapidly than sensitivity, suggesting that the classifier initially favors the dominant or easier class, effectively learning to reject the minority composite class but failing to identify it correctly.

Around the intermediate training regime, approximately between 400 and 600 sweeps, a sharp transition occurs. Test sensitivity, which had remained suppressed up to that point, suddenly increases and rapidly aligns with the training curve, while overall test accuracy exhibits a concurrent jump. This abrupt improvement in generalization, without a corresponding discontinuity in training performance, is the hallmark of grokking: the model reorganizes its internal representation from a memorization-dominated regime to one that captures the underlying structure of the classification problem. In the context of a matrix product state architecture, this transition can be interpreted as the point at which the network develops sufficient effective “entanglement” or correlation capacity to encode global patterns rather than relying on local heuristics.

After this transition, both training and test metrics stabilize at high values with a small generalization gap, indicating that the learned representation is expressive and well-regularized. The convergence of sensitivity and specificity further confirms that the classifier has achieved a balanced discrimination capability, no longer privileging one class at the expense of the other. This behavior is consistent with a compression phase in which the MPS reorganizes into a lower-complexity, structured representation that supports robust generalization.
The multiclass dynamics shown in the right panel exhibit a more nuanced version of the same phenomenon. Instead of a single sharp transition, the grokking process appears distributed across classes and unfolds over a broader range of training sweeps. Early in training, all class-wise test accuracies are low and unstable, with some curves even decreasing, suggesting an initial regime of representation collapse or confusion where the model has not yet established a meaningful feature space. As training proceeds, performance improves rapidly but unevenly across classes: class 0 achieves high accuracy relatively quickly, approaching values near $80 \, \%$, while classes 1 2 and 3 plateau at substantially lower levels, typically between $40 \, \%$ and $60 \, \%$.

This stratification reflects intrinsic complexity differences. The dominant class, which is more homogeneous and better represented, is learned early and robustly, effectively driving the initial gains in overall accuracy. In contrast classes 1, 2, and 3 likely require modeling more subtle, nonlinear interactions among features and therefore benefit less from the initial stages of representation learning. The absence of a sharp, simultaneous jump across all classes indicates that the grokking transition in the multiclass setting is hierarchical: the model first internalizes the coarse structure that separates the dominant class from the rest, and only later refines its representation to distinguish among the minority classes.
Overall, the observed behavior highlights how grokking in MPS-based classifiers depends strongly on task structure. In simpler, binary settings, the emergence of generalization manifests as a clear phase transition, whereas in more complex, multiclass problems it becomes fragmented and class-dependent, reflecting the progressive acquisition of increasingly fine-grained decision boundaries.

\subsection*{Magnetization patterns}

\begin{figure*}[t!]
\includegraphics[width=\linewidth]{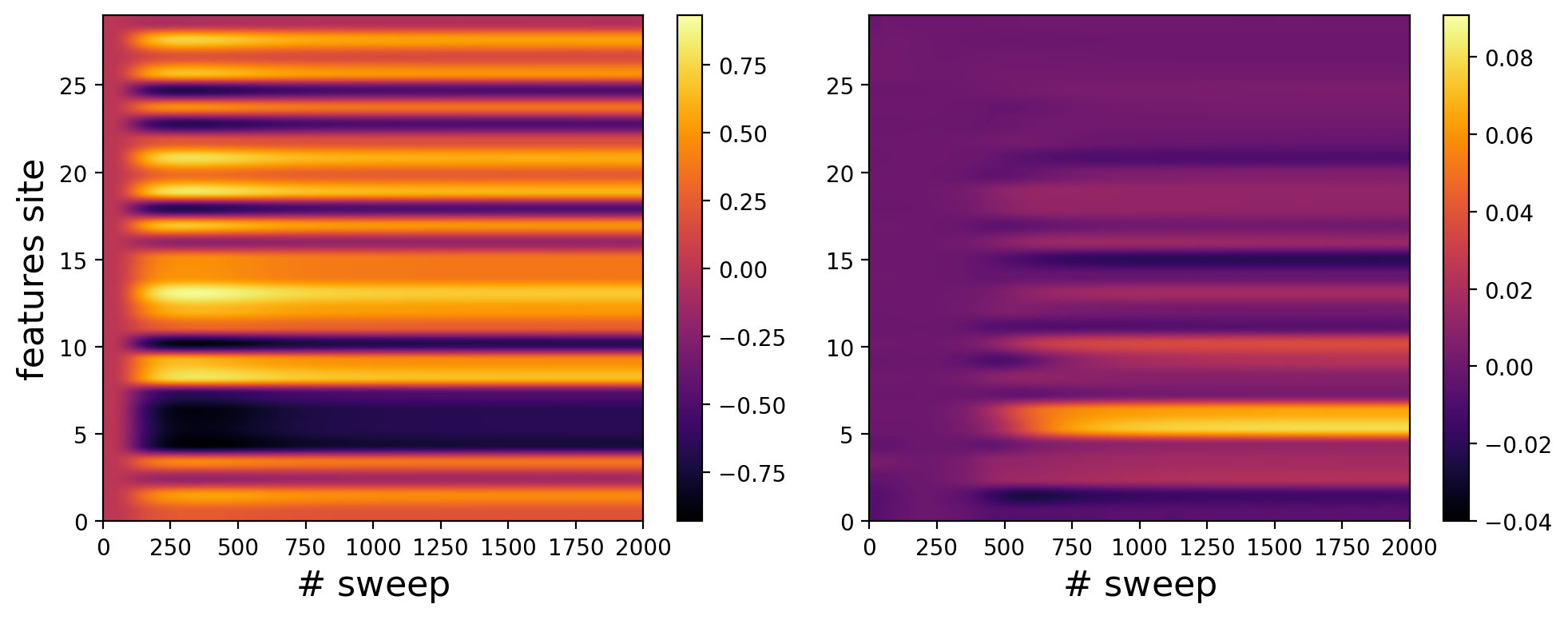}
\caption{Magnetization dynamics of the MPS quantum mask for the binary classification problem, class 0 (left panel) vs classes 1–3 (right panel), showing the evolution of Pauli $\sigma_Z$ expectation values across feature sites and training sweeps. Average magnetization is computed over three independent runs; the standard deviation exhibits a single peak of approximately $7 \times 10^{-2}$, observed only in the class‑0 mask on the left near the grokking transition, followed by a rapid decay toward negligible values. \label{fig:magnetization_binary}}
\end{figure*}

The internal dynamics of the MPS quantum mask during training can be explicitly monitored through the expectation values of the Pauli operator $\sigma_Z$ at each feature site, providing a physically interpretable proxy for feature activation across training sweeps. These magnetization patterns offer a direct view of how the model progressively organizes information in the embedding space, revealing both the emergence of structured correlations and the stabilization associated with generalization.

In the binary classification setting, the magnetization map shown in Figure \ref{fig:magnetization_binary} exhibits a well-defined stratified structure across feature sites. After an initial transient phase, clear horizontal bands emerge and stabilize over the sweeps, reflecting the progressive alignment of specific features with the learned decision boundary. The magnitude and sign of the $\sigma_Z$ expectation values indicate that the MPS assigns consistent roles to subsets of features, effectively encoding the separation between class and the aggregated 1–3 classes. This structured pattern coincides temporally with the grokking transition observed in classification performance, suggesting that the abrupt gain in generalization is accompanied by a global reorganization of the quantum mask into a coherent, low-complexity configuration.

\begin{figure*}[t!]
\includegraphics[width=\linewidth]{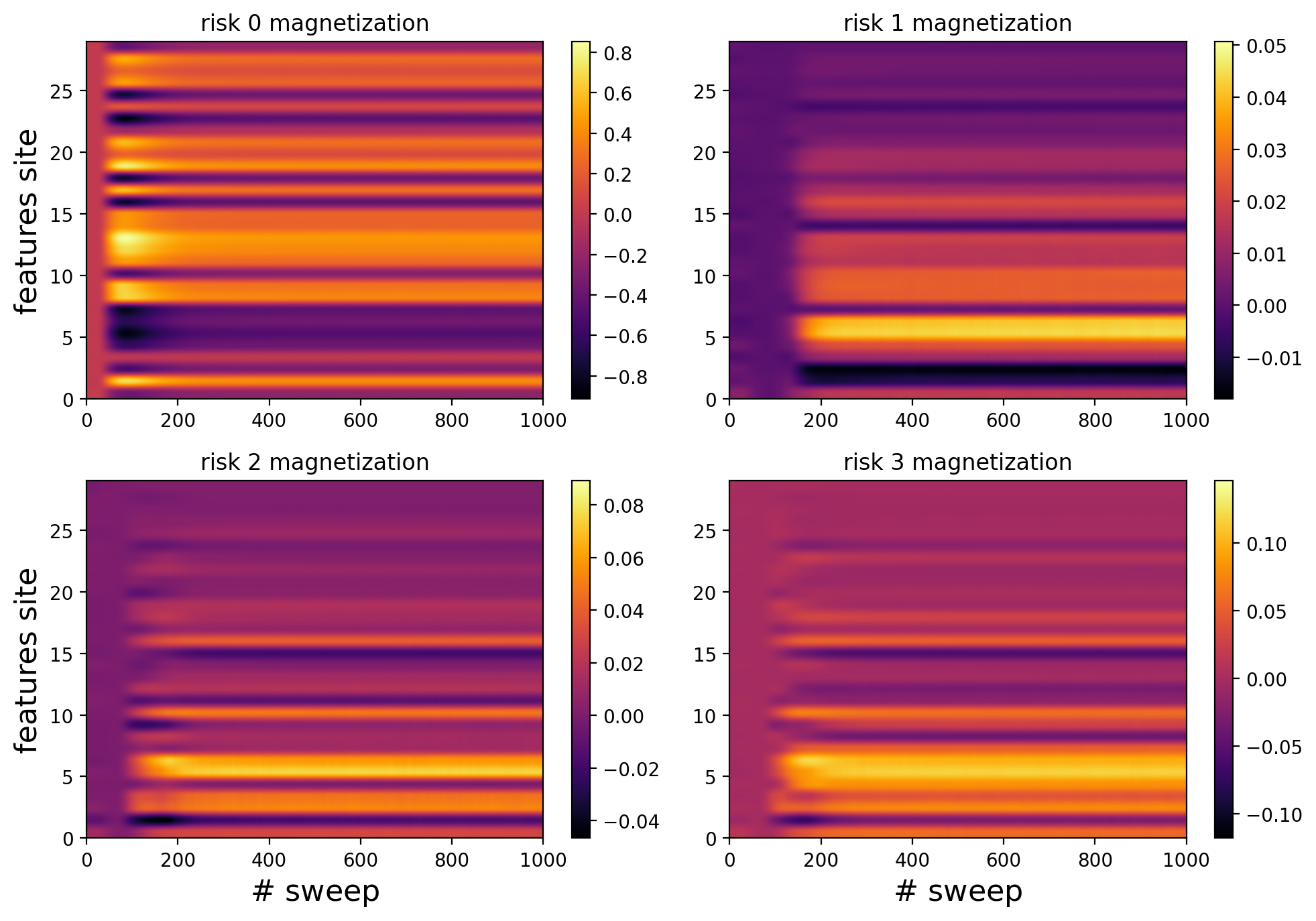}
\caption{Class-wise magnetization patterns for the four-class classification problem, illustrating the evolution of $\sigma_Z$ expectation values for each susceptibility level. Average magnetization is computed over three independent runs; the standard deviation exhibits a single peak of approximately $8 \times 10^{-2}$, observed only in the class‑0 mask near the grokking transition, followed by a rapid decay toward negligible values.
\label{fig:magnetization_4class}}
\end{figure*}

A remarkable consistency emerges when comparing these patterns with those obtained in the multiclass setting in Figure \ref{fig:magnetization_4class}. The magnetization associated with class 0 in the four-class problem reproduces almost perfectly the structure observed in the binary case for class 0, indicating that the representation learned for the dominant low-susceptibility class is stable and largely independent of the classification formulation. Conversely, the pattern associated with the aggregated classes 1-3 in the binary case closely matches the individual magnetization maps obtained for classes 1, 2, and 3 in the multiclass configuration. Although the amplitudes are lower and the patterns appear more diffuse for individual classes, the spatial arrangement of activated feature sites remains consistent, demonstrating that the model captures a shared underlying structure for higher vulnerability conditions.

This correspondence highlights an important property of the MPS-based classifier: the quantum mask learns hierarchical and compositional representations, where the binary separation emerges as a coarse-grained projection of the more detailed multiclass structure. The near-perfect alignment of magnetization patterns across formulations confirms that the grokking transition is not merely a performance artifact, but reflects the stabilization of a physically meaningful internal representation, in which feature sites acquire stable “magnetic” roles encoding wildfire susceptibility information.

\subsection*{Quantum masks mixedness}

\begin{figure*}[t!]
\includegraphics[width=\linewidth]{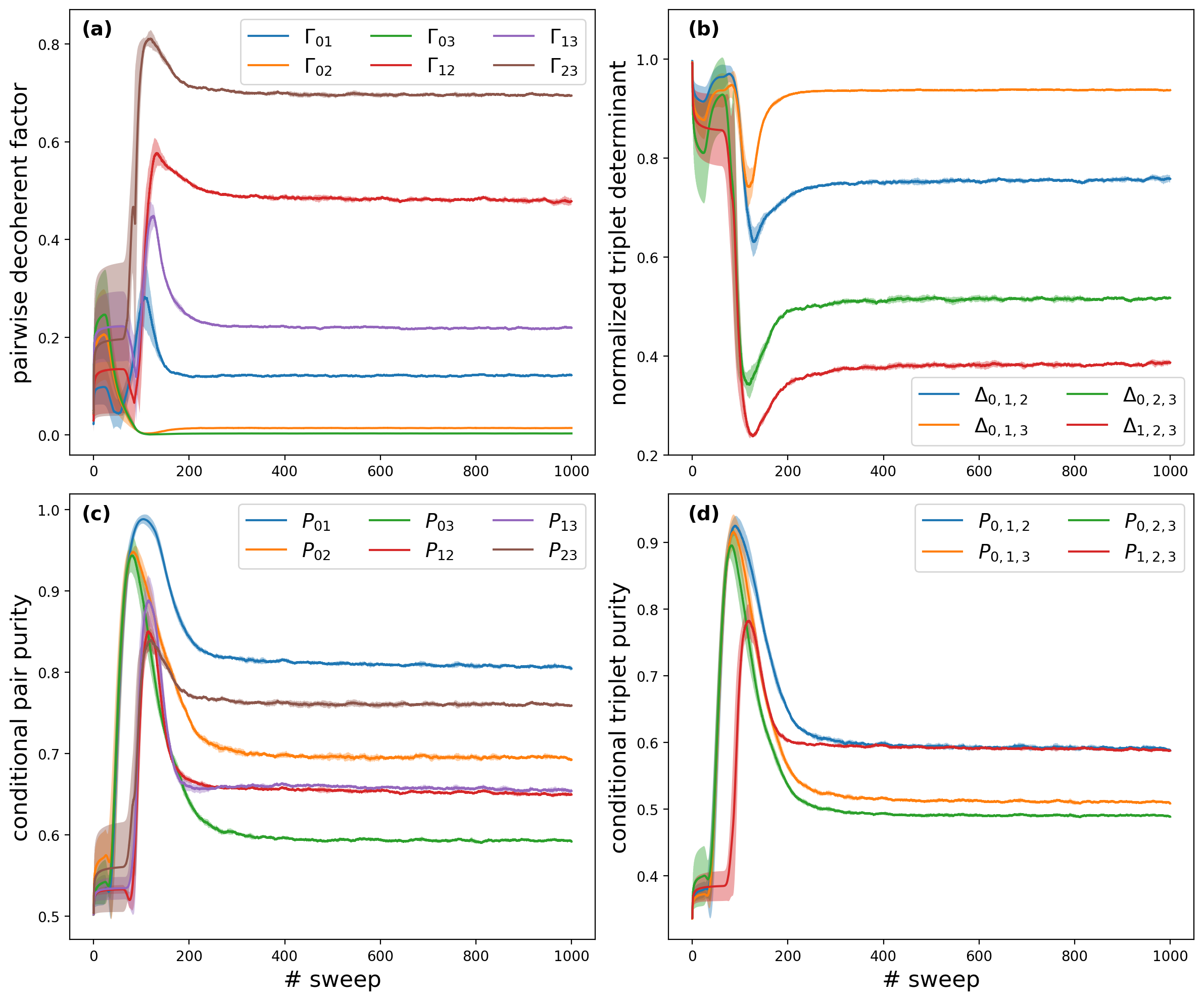}
\caption{Evolution during four-class classification training of level-resolved mixedness metrics during MPS training. Panels (a–d) report pairwise decoherence factors $\Gamma_{ij}$, normalized triplet determinants $\Delta_{ijk}$, and conditional purities for pairs and triplets of susceptibility levels. Solid lines denote the mean values, while shaded regions represent the corresponding standard deviation computed over three independent runs. \label{fig:mixedness}}
\end{figure*}

The mixedness properties of the quantum masks provide a complementary and deeper characterization of the learning dynamics, beyond the magnetization patterns, by explicitly quantifying the degree of distinguishability encoded in the label reduced density matrices. The level-resolved quantities introduced in Methods section, namely the pairwise decoherence factors $\Gamma_{ij}$, the normalized triplet determinants $\Delta_{ijk}$, and the corresponding conditional purities, allow us to track how information about the different wildfire vulnerability levels is progressively encoded in the environment during training in Figure \ref{fig:mixedness}.

Panel (a) shows the evolution of the pairwise decoherence factors $\Gamma_{ij}$, which quantify the residual coherence between pairs of levels. A striking structure emerges after the initial transient: pairs corresponding to adjacent categories, in particular $(1,2)$ and $(2,3)$, systematically retain higher values of $\Gamma_{ij}$, indicating stronger coherence and therefore reduced distinguishability leakage to the environment. In contrast, non-adjacent pairs such as $(0,2)$ or $(0,3)$ display markedly lower values, signaling stronger decoherence and thus a higher degree of environmental distinguishability, increasingly mixed with label states. This asymmetry reflects the fact that the model separates well distant susceptibility categories, while neighboring classes remain partially overlapping in feature space.

This behavior is further corroborated in panel (b), where the normalized triplet determinants $\Delta_{ijk}$ capture higher-order correlations among triplets of levels. Triplets that include well-separated categories, such as $(0,1,3)$, stabilize at significantly higher values compared to those involving only adjacent classes, such as $(1,2,3)$. Since $\Delta_{ijk}$ quantifies the volume spanned by the corresponding conditional states, larger values indicate that the environment encodes more independent information about these levels, thereby enhancing their distinguishability. The consistent ordering observed across triplets confirms that non-adjacent classes remain more geometrically separated in the learned representation.

Panels (c) and (d) provide an alternative perspective through the conditional purities of pairwise and triplet subspaces. Here again, subsets involving adjacent vulnerability levels systematically retain higher purity, meaning that lower environment mixedness is induced in these subspaces during training. In particular, pairs such as $(0,1)$, $(2,3)$ and triplets such as $(0,1,2)$, $(1,2,3)$ converge to higher purity plateaus compared to subsets composed of non neighboring classes. This indicates that the corresponding subspaces preserve a larger degree of internal coherence, consistent with increased overlap between their associated conditional states.

Altogether, all four diagnostics consistently reveal the same fundamental structure: non-adjacent  categories are more sharply distinguished by the quantum mask than adjacent ones. This pattern emerges robustly across pairwise, triplet, and normalized measures, and persists after the grokking transition, indicating that it is an intrinsic property of the learned representation rather than a transient training artifact. From a physical perspective, this means that the environment acquires more independent information for separating distant classes, while only partial information is available to resolve finer distinctions between neighboring levels. In turn, this provides a principled explanation for the classification behavior observed in the multiclass setting, where confusion is predominantly localized among adjacent susceptibility categories.

\section*{Discussion}

%\textcolor{blue}{The Discussion should be succinct and must not contain subheadings.}

Wildfire susceptibility mapping remains a challenging task because fire occurrence is governed by the interaction of climatic, ecological, geomorphological, and anthropogenic factors, rather than by a single dominant control. In Mediterranean landscapes, vegetation structure, fuel continuity, climatic stress, land-use patterns, and human pressure jointly shape ignition probability and fire-prone conditions. This complexity motivates the use of machine-learning approaches capable of capturing non-linear relationships within heterogeneous spatial data.
The relevance of this problem extends beyond wildfire occurrence alone. Wildfires can act as triggering events within broader hazard cascades, altering vegetation cover, runoff generation, soil erosion, sediment connectivity, and slope stability \cite{SHAKESBY2006,MOODY2013,ZINGARO2024}. Accurate wildfire susceptibility classification therefore provides useful information not only for fire-risk assessment, but also for anticipating secondary geomorphological and hydrological impacts in multi-hazard monitoring frameworks \cite{gill2014,gill2016}.
Within this context, AlphaEarth embeddings offer a scalable alternative to classical susceptibility modelling approaches based on manually harmonized predictor layers. By providing a compact representation learned from multi-source Earth observation data, embeddings reduce the burden of dataset preprocessing while retaining information relevant to complex environmental gradients. This makes them particularly suitable for testing whether learned geospatial representations can support wildfire susceptibility classification in heterogeneous Mediterranean regions. In this context, the MPS classifier is useful not simply as a predictive model, but as a way to probe the learnability of wildfire susceptibility classes from AlphaEarth embeddings.
The classical wildfire classification approach for the Gargano region provides a strong benchmark, with the Random Forest model achieving an overall accuracy above $95\%$ and consistently high precision, recall, and F1 score \cite{wildfire2026}. This performance indicates that wildfire-prone areas can be effectively identified using explicit environmental predictors such as fuel load, land cover, and climatic variables.
While the Random Forest framework reported in the literature achieves near-optimal performance for wildfire susceptibility mapping in the Gargano  region, the results obtained with the MPS classifier provide complementary insights that extend beyond predictive accuracy. In particular, the training dynamics reveal the emergence of a grokking transition in the binary formulation, where the separation between low-susceptibility areas (class 0) and the aggregated higher-susceptibility classes (1–3) undergoes a delayed but abrupt improvement in generalization. This transition is accompanied by the spontaneous stabilization of the quantum-mask magnetization patterns, indicating the formation of a coherent internal feature representation. Moreover, the class-resolved magnetization maps obtained in the multiclass setting closely reproduce the structures observed in the binary problem: the pattern associated with class 0 is nearly indistinguishable from the binary class 0 mask, while the patterns associated with classes 1, 2, and 3 collectively resemble the higher-susceptibility aggregate class identified in the binary formulation. This strong consistency suggests that the MPS architecture learns a hierarchical organization of wildfire susceptibility, where multiclass discrimination emerges as a refinement of a more fundamental low-susceptibility/high-susceptibility separation.

An even more distinctive aspect of the proposed framework emerges from the analysis of reduced density matrices and level-resolved mixedness measures. Unlike conventional machine-learning approaches, which typically provide only classification scores and feature importance rankings, the tensor-network representation allows direct access to physically interpretable quantities related to coherence and distinguishability among classes. The pairwise decoherence factors, normalized triplet determinants, and subspace purities consistently reveal that non-adjacent susceptibility categories are more easily distinguishable than neighboring classes. In other words, the learned representation naturally separates low-vulnerability and high-vulnerability conditions, whereas confusion remains concentrated among adjacent susceptibility levels. This observation is fully consistent with the class-wise classification results, where errors predominantly occur between neighboring categories rather than between extreme classes. From an operational perspective, this hierarchy of distinguishability provides valuable information about the intrinsic structure of wildfire susceptibility labels and complements traditional XAI analyses by offering a quantitative characterization of inter-class relationships. Therefore, although the MPS classifier does not currently reach the exceptionally high accuracies reported for Random Forest models, it provides a richer description of the learning process and reveals latent geometric properties of the wildfire susceptibility representation that remain inaccessible to standard supervised learning methods \cite{pomarico2025grokkingentanglementtransitiontensor, technologies13100438}.

\section*{Methods}

\subsection*{MPS classifier}

Tensor network approaches to machine learning exploit the remarkable ability of these structures to encode and manipulate high-dimensional information in a compressed and structured form. Among the various architectures, MPS, also known in numerical analysis as tensor trains, represent one of the most prominent and versatile paradigms~\cite{Tucker1966, hackbusch2019, delathauwer2000, oseledets2011}. Their defining feature is the factorization of high-order tensors into a sequence of interconnected low-rank components, drastically reducing the number of free parameters while retaining a rich expressive capacity.

Originally developed within the context of quantum many-body theory and quantum information science, tensor networks have become indispensable tools for describing both equilibrium states and dynamical processes in complex quantum systems~\cite{PhysRevLett.69.2863, RevModPhys.77.259, SCHOLLWOCK201196}. A particularly appealing aspect of MPS representations is the tunable control over their complexity. This is achieved through the bond dimension $\chi$, which determines the number of retained singular components in the decomposition and thus regulates the trade-off between model flexibility and computational efficiency. By adjusting $\chi$, one can systematically interpolate between simple, low-rank approximations and highly expressive representations, making MPS architectures well suited for scalable machine learning applications. Combined with their inherent interpretability, these properties position tensor networks as a natural bridge between quantum-inspired models and classical learning frameworks.

We adopt a quantum-inspired encoding scheme in which classical input vectors are lifted into a high-dimensional Hilbert space, enabling nonlinear feature representations through linear operations in the embedded space. Specifically, each input vector $\mathbf{x} \in \mathbb{R}^N$ is mapped to a tensor product state
$\ket{\Phi(\mathbf{x})} = \bigotimes_{j=1}^{N} \ket{\phi(x^{(j)})}$,
where each scalar component is embedded into a two-dimensional local state $\ket{\phi(x^{(j)})} = (\cos x^{(j)}, \sin x^{(j)})^\intercal$. This construction preserves certain geometric relationships of the original data while enabling the model to capture nonlinear dependencies through contractions in the enlarged feature space~\cite{stoudenmire2017supervisedlearningquantuminspiredtensor}.

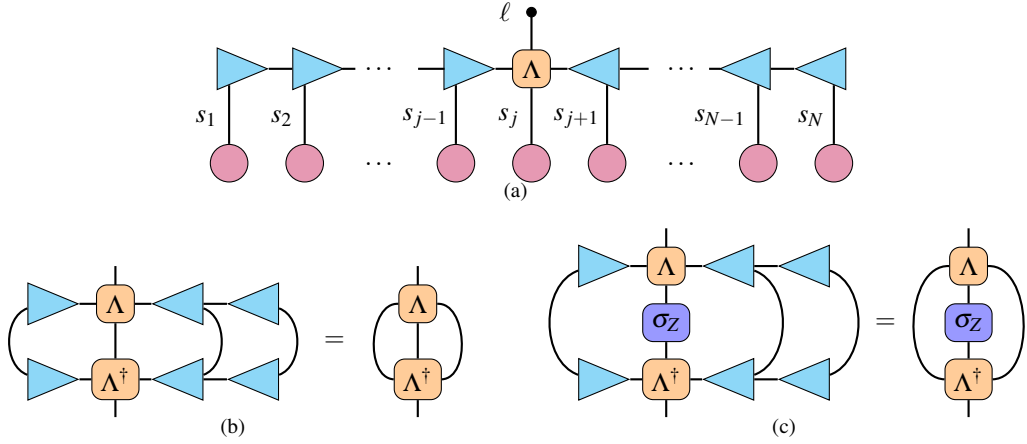
\begin{figure*}[t!]
\begin{center}
\begin{tabular}{c}
    \subfigure[]{\centering
    \begin{tikzpicture}
    \node[shape=isosceles triangle, draw, fill=cyan!40, minimum size=0.55cm] (v0z) at (-1,0.5) {};
    \node[shape=isosceles triangle, draw, fill=cyan!40, minimum size=0.55cm] (v0b) at (0,0.5) {};
    \node[shape=isosceles triangle, draw, fill=cyan!40, minimum size=0.55cm] (v0a) at (2,0.5) {};
    \node[draw, rounded corners, fill=orange!40, shape=rectangle, minimum width=0.5cm, minimum height = 0.5cm] (v0c) at (3,0.5) {$\Lambda$};
    %\node[draw, fill=red!40, shape=rectangle, minimum width=0.5cm, minimum height = 0.5cm] (v1) at (4,0.5) {};
    %\node[draw, fill=red!40, shape=rectangle, minimum width=0.5cm, minimum height = 0.5cm] (v2) at (5,0.5) {};
    \node[shape=isosceles triangle, draw, fill=cyan!40, minimum size=0.55cm, rotate=180] (v2a) at (4,0.5) {};
    \node[shape=isosceles triangle, draw, fill=cyan!40, minimum size=0.55cm, rotate=180] (v2b) at (6,0.5) {};
    \node[shape=isosceles triangle, draw, fill=cyan!40, minimum size=0.55cm, rotate=180] (v2z) at (7,0.5) {};
    \node[draw, fill=purple!40, shape=circle, inner sep=5 pt] (v00z) at (-1,-0.75) {};
    \node[draw, fill=purple!40, shape=circle, inner sep=5 pt] (v00) at (0,-0.75) {};
    \node[draw, fill=purple!40, shape=circle, inner sep=5 pt] (v00a) at (2,-0.75) {};
    \node[draw, fill=purple!40, shape=circle, inner sep=5 pt] (v00b) at (3,-0.75) {};
    %\node[draw, fill=blue!40, shape=circle, inner sep=5 pt] (v22b) at (5,-0.75) {};
    \node[draw, fill=purple!40, shape=circle, inner sep=5 pt] (v22a) at (4,-0.75) {};
    \node[draw, fill=purple!40, shape=circle, inner sep=5 pt] (v22c) at (6,-0.75) {};
    \node[draw, fill=purple!40, shape=circle, inner sep=5 pt] (v22z) at (7,-0.75) {};
    \draw [thick] (v0b) -- (v00)
    (v0c) -- (v00b)
    (v0c) -- (v2a)
    (v0c) -- (v0a)
    (v2a) -- (v22a)
    (v0a) -- (v00a)
    (v0z) -- (v00z)
    (v0b) -- (v0z)
    %(v2) -- (v22b)
    (v2b) -- (v2z)
    (v2z) -- (v22z)
    (v2b) -- (v22c);
    \draw [thick] (3,0.75) -- (3,1.25);
    %\draw [thick] (5,0.25) -- (5,-0.25);
    %\draw [thick] (7,0.25) -- (7,-0.25);
    \draw [thick] (0.67,0.5) -- (0.5,0.5);
    \draw [thick] (1.5,0.5) -- (1.85,0.5);
    \draw [thick] (4.17,0.5) -- (4.55,0.5);
    \draw [thick] (5.5,0.5) -- (5.35,0.5);
    \node [] (t) at (5,0.5) {$\dots$};
    \node [] (td) at (5,-0.75) {$\dots$};
    \node [] (tb) at (1,0.5) {$\dots$};
    \node [] (tbd) at (1,-0.75) {$\dots$};
    \draw[black,fill=black] (3,1.25) circle (.4ex);
    \node [] (t3) at (2.65,1.25) {$\ell$};
    \node [] (t2) at (-1.3,-0.15) {$s_1$};
    \node [] (t2) at (-0.3,-0.15) {$s_2$};
    \node [] (t2) at (1.6,-0.15) {$s_{j-1}$};
    \node [] (t2) at (2.7,-0.15) {$s_j$};
    \node [] (t2) at (3.6,-0.15) {$s_{j+1}$};
    %\node [] (t5) at (4.5,-0.15) {$s_{\ell+1}$};
    \node [] (t6) at (5.5,-0.15) {$s_{N-1}$};
    \node [] (t6) at (6.7,-0.15) {$s_N$};
    \end{tikzpicture}} \\
    \subfigure[]{\centering
    \begin{tikzpicture}
    \node[shape=isosceles triangle, draw, fill=cyan!40, minimum size=0.55cm] (v0b) at (2,0.5) {};
    \node[draw, rounded corners, fill=orange!40, shape=rectangle, minimum width=0.5cm, minimum height = 0.5cm] (v0c) at (3,0.5) {$\Lambda$};
    \node[shape=isosceles triangle, draw, fill=cyan!40, minimum size=0.55cm, rotate=180] (v1) at (4,0.5) {};
     \node[shape=isosceles triangle, draw, fill=cyan!40, minimum size=0.55cm, rotate=180] (v2b) at (5,0.5) {};
    \node[shape=isosceles triangle, draw, fill=cyan!40, minimum size=0.55cm] (v00b) at (2,-0.5) {};
    \node[draw, rounded corners, fill=orange!40, shape=rectangle, minimum width=0.5cm, minimum height = 0.5cm] (v00c) at (3,-0.5) {$\Lambda^\dagger$};
    \node[shape=isosceles triangle, draw, fill=cyan!40, minimum size=0.55cm, rotate=180] (v11) at (4,-0.5) {};
     \node[shape=isosceles triangle, draw, fill=cyan!40, minimum size=0.55cm, rotate=180] (v22b) at (5,-0.5) {};
    \draw [thick] (v0b) -- (v0c)
    (v0c) -- (v1)
    (v1) -- (v2b)
    %(v0b) -- (v00b)
    (v0c) -- (v00c)
    %(v1) -- (v11)
    %(v2b) -- (v22b)
    (v00b) -- (v00c)
    (v00c) -- (v11)
    (v11) -- (v22b);
    \draw [thick] (3,0.75) -- (3,1);
    \draw [thick] (3,-0.775) -- (3,-1);
    \node [] (t2) at (5.9,0) {$=$};
     \node[draw, rounded corners, fill=orange!40, shape=rectangle, minimum width=0.5cm, minimum height = 0.5cm] (v0d) at (7,0.5) {$\Lambda$};
    \node[draw, rounded corners, fill=orange!40, shape=rectangle, minimum width=0.5cm, minimum height = 0.5cm] (v00d) at (7,-0.5) {$\Lambda^\dagger$};
    \draw [thick] (v0b) to [bend left=-70] (v00b);
    \draw [thick] (v1) to [bend left=70] (v11);
    \draw [thick] (v2b) to [bend left=70] (v22b);
    \draw [thick] (v0d) -- (v00d);
    \draw [thick] (v0d) to [bend left=90] (v00d);
    \draw [thick] (v0d) to [bend right=90] (v00d);
    \draw [thick] (7,0.75) -- (7,1);
    \draw [thick] (7,-0.775) -- (7,-1);
    %\node [] (t3) at (9.35,0) {\( \displaystyle = \varrho^{(\ell)} = \begin{pmatrix}
    %    \varrho^{(\ell)}_{00} & \varrho^{(\ell)}_{01} \\
    %    \varrho^{(\ell)}_{10} & \varrho^{(\ell)}_{11}
    %\end{pmatrix} \)};
    \end{tikzpicture}} $\qquad$
    \subfigure[]{\centering
    \begin{tikzpicture}
    \node[shape=isosceles triangle, draw, fill=cyan!40, minimum size=0.55cm] (v0b) at (2,0.5) {};
    \node[draw, rounded corners, fill=orange!40, shape=rectangle, minimum width=0.5cm, minimum height = 0.5cm] (v0c) at (3,0.5) {$\Lambda$};
    \node[shape=isosceles triangle, draw, fill=cyan!40, minimum size=0.55cm, rotate=180] (v1) at (4,0.5) {};
     \node[shape=isosceles triangle, draw, fill=cyan!40, minimum size=0.55cm, rotate=180] (v2b) at (5,0.5) {};
    \node[shape=isosceles triangle, draw, fill=cyan!40, minimum size=0.55cm] (v00b) at (2,-1) {};
    \node[draw, rounded corners, fill=orange!40, shape=rectangle, minimum width=0.5cm, minimum height = 0.5cm] (v00c) at (3,-1) {$\Lambda^\dagger$};
    \node[shape=isosceles triangle, draw, fill=cyan!40, minimum size=0.55cm, rotate=180] (v11) at (4,-1) {};
    \node[shape=isosceles triangle, draw,fill=cyan!40, minimum size=0.55cm, rotate=180] (v22b) at (5,-1) {};
    \node[draw, rounded corners, fill=blue!40, shape=rectangle, minimum width=0.5cm, minimum height = 0.5cm] (vint) at (3,-0.25) {$\sigma_Z$};
    \draw [thick] (v0b) -- (v0c)
    (v0c) -- (v1)
    (v1) -- (v2b)
    %(v0b) -- (v00b)
    %(v1) -- (v11)
    (v0c) -- (vint)
    (vint) -- (v00c)
    %(v2b) -- (v22b)
    (v00b) -- (v00c)
    (v00c) -- (v11)
    (v11) -- (v22b);
    \draw [thick] (3,0.75) -- (3,1);
    \draw [thick] (3,-1.275) -- (3,-1.5);
    \node [] (t2) at (5.9,-0.25) {$=$};
    \node[draw, rounded corners, fill=orange!40, shape=rectangle, minimum width=0.5cm, minimum height = 0.5cm] (v0d) at (7,0.5) {$\Lambda$};
    \node[draw, rounded corners, fill=orange!40, shape=rectangle, minimum width=0.5cm, minimum height = 0.5cm] (v00d) at (7,-1) {$\Lambda^\dagger$};
    \node[draw, rounded corners, fill=blue!40, shape=rectangle, minimum width=0.5cm, minimum height = 0.5cm] (vintb) at (7,-0.25) {$\sigma_Z$};
    \draw [thick] (v0b) to [bend left=-70] (v00b);
    \draw [thick] (v1) to [bend left=70] (v11);
    \draw [thick] (v2b) to [bend left=70] (v22b);
    \draw [thick] (v0d) to [bend right=90] (v00d)
    (v0d) -- (vintb)
    (vintb) -- (v00d)
    (v0d) to [bend left=90] (v00d);
    \draw [thick] (7,0.75) -- (7,1);
    \draw [thick] (7,-1.275) -- (7,-1.5);
    \end{tikzpicture}} \\
\end{tabular}
\end{center}
\caption{Diagrammatics for MPS classifier quantities. The~qubit encoding is represented with pink circles, in contraction to define the predictor shown in panel (a), where the variational tensor $W$ is expressed as a MPS in the mixed-canonical form. The exploitation of an orthogonality center $\Lambda$ enables efficient tensor contractions, for the pictorial purpose of a $N=4$ lattice; (b) the reduced density matrix in label space, and~(c) the local magnetization at each site. \label{fig::mps}}
\end{figure*}

The classification model is defined through a variational tensor object, referred to as a quantum mask $W$, which acts on the encoded state, as depicted in Figure \ref{fig::mps}(a). To ensure computational feasibility, $W$ is expressed in MPS form, allowing efficient manipulation of exponentially large tensors through a chain decomposition controlled by the bond dimension $\chi$. The construction relies on successive singular value decompositions (SVD) at each bipartition, where only the leading $\chi$ singular values are retained, effectively limiting the growth of correlations while maintaining the dominant contributions. Within this framework, the MPS represents a class dependent tensor network whose contraction with the input state produces a set of prediction scores $f_W^\ell(\mathbf{x})= \sum_{s_1 \dots s_N} W^\ell_{s_1,\dots,s_N} \ket{\phi(x^{(1)})^{s_1}} \otimes \dots \otimes \ket{\phi(x^{(N)})^{s_N}}$, one for each label $\ell$. The predicted class corresponds to the component with the largest amplitude. Model parameters are learned by minimizing a mean squared error loss $\mathcal{C}(W) = \frac{1}{2} \sum_{\omega} \sum_\ell \left( f^\ell_{W}(\mathbf{x}_{\omega}) - y^\ell_{\omega} \right)^2$ between predicted amplitudes and target label encodings $y^\ell_{\omega}$ across the training dataset, with $\omega = 1, \dots, N_T$. This formulation differs from probabilistic approaches based on Born-rule normalization, instead operating directly on amplitude-based scores, following the conventions adopted in prior tensor network learning approaches.

Optimization is carried out using a two-site update strategy, in which pairs of adjacent tensors are jointly optimized at each step. This procedure involves constructing an effective two-site tensor, performing a gradient-based update, and subsequently decomposing it via truncated SVD to restore the MPS structure. Sweeping this update along the chain allows for iterative refinement of the model parameters. Because gradient descent induces non-unitary evolution of the tensor network, a renormalization step is required after each update to maintain numerical stability and meaningful normalization. This is implemented through bond dimension truncation and global rescaling, effectively projecting the updated state onto its most relevant subspace, analogous to a measurement process selecting dominant components~\cite{pomarico2025grokkingentanglementtransitiontensor,technologies13100438,wiersema2023}. To improve both computational efficiency and scalability, the optimization of the tensor $B$ is performed using a stochastic gradient descent scheme. Instead of evaluating the gradient over the entire training dataset at each step, we approximate it by sampling mini-batches of input examples, thus significantly reducing computational overhead while preserving convergence properties. Concretely, the gradient update is computed over randomly selected balanced subsets of the training data, enabling an efficient exploration of the loss landscape and introducing a controlled level of stochasticity that can help avoid poor local minima. All training procedures are executed on a high-performance GPU architecture, specifically an NVIDIA H100 equipped with 40 GB of memory. This hardware allows for efficient tensor contractions, parallel batch processing, and rapid execution of the repeated sweep updates inherent to the MPS optimization algorithm, thereby enabling the training of large-scale models and the analysis of their dynamical properties within feasible computational time. The batch size is equal to 512 for both classification problems, bond dimension $\chi=1000$ and learning rate $\alpha \approx 5 \times 10^{-3}$ for the binary case, while for the multiclass problem $\chi=500$ and $\alpha \approx 7 \times 10^{-4}$. The number of 30 features was selected after benchmarking the binary classification task by sampling $10^3$ embedding pixels from each training super-pixel, using 20, 30, and 40 features, which yielded test accuracies of $79.8\%$, $83.3\%$, and $82.9\%$, respectively, indicating optimal performance for the intermediate feature set.

To gain insight into the internal learning dynamics, we monitor physically motivated observables derived from the tensor representation. In particular, we evaluate reduced density matrices in the label space, as shown in Figure \ref{fig::mps}(b), which encode coherence and distinguishability among classes, as well as local expectation values of Pauli operators at each feature site, represented  in Figure \ref{fig::mps}(c). These quantities provide an interpretable picture of how individual features contribute to the classification decision and how class-specific information is distributed across the network.

%\textcolor{blue}{Topical subheadings are allowed. Authors must ensure that their Methods section includes adequate experimental and characterization data necessary for others in the field to reproduce their work.}

\subsection*{Level-resolved mixedness measures for reduced density matrices}

Consider a bipartite quantum system $A\otimes B$ prepared in a pure state $\ket{\psi}= \sum_{i=0}^{d-1}
|i\rangle_A |\phi_i\rangle_B$, where ${|i\rangle}$ denotes an orthonormal basis of the $d$-dimensional subsystem $A$, while the vectors $|\phi_i\rangle$ belong to the Hilbert space of subsystem $B$. Tracing out subsystem $B$ yields the reduced density matrix $\rho_A = \mathrm{Tr}_B(\ket{\psi}\bra{\psi})=\sum_{ij}
\braket{\phi_j|\phi_i}
\ket{i}\bra{j}$, with matrix elements $\rho_{ii} = |\phi_i|^2, \rho_{ij} =
\braket{\phi_j|\phi_i}, i\neq j$. The diagonal entries therefore represent the populations of the different qudit levels, while the off-diagonal entries quantify the overlap between the corresponding conditional states of subsystem $B$. Whenever $\rho_{ii}=0$, the corresponding basis state never appears in the global wavefunction and therefore does not participate in the dynamics. According to Sylvester's criterion \cite{horn2012matrix}, since every density matrix is positive semidefinite, every $2\times2$ principal minor satisfies $\rho_{ii}\rho_{jj}
\ge |\rho_{ij}|^2$, which is simply the Cauchy-Schwarz inequality applied to the vectors
$\ket{\phi_i}$ and $\ket{\phi_j}$. Equality, $\rho_{ii}\rho_{jj}=|\rho_{ij}|^2$, holds if and only if $\ket{\phi_i}
\propto \ket{\phi_j}$, meaning that subsystem $B$ cannot distinguish whether subsystem $A$ occupies level $\ket{i}$ or $\ket{j}$. Therefore $\ket{\psi}=(\alpha \ket{i}_A + \beta \ket{j}_A) \otimes \ket{\phi}_B + \dots$ inside the subspace, hence the pair
$\{ \ket{i},\ket{j} \}$ forms a pure local superposition, yielding maximal coherence. In contrast, strict inequality indicates that information about these alternatives has leaked into subsystem $B$, leading to a reduction of local coherence, that is the environment possesses information capable of distinguishing the two levels and the coherence between levels $i$ and $j$ is reduced because the corresponding amplitudes are entangled with distinguishable states of subsystem $B$. This observation motivates a hierarchy of level-resolved purity diagnostics.

\subsubsection*{Pairwise determinant and decoherence factor}

For every pair of levels $(i,j)$ we define the determinant of the corresponding
$2\times2$ principal minor,
$D_{ij} =\rho_{ii}\rho_{jj}
-|\rho_{ij}|^2$. It is known as Gram determinant and positivity of $\rho$ guarantees $D_{ij} \ge 0$ \cite{horn2012matrix}. This quantity provides a direct measure of the loss of local coherence between two levels, since
\begin{itemize}
\item $D_{ij}=0$ if and only if the two-level block has rank one and therefore represents a locally pure superposition;
\item $D_{ij}>0$ indicates that the two-level sector is mixed, implying that subsystem $B$ carries information capable of distinguishing the two alternatives.
\end{itemize}
A particularly intuitive quantity is obtained by normalizing the coherence with respect to its maximal allowed value,
\begin{equation}
\Gamma_{ij} = \frac{|\rho_{ij}|}{\sqrt{\rho_{ii}\rho_{jj}}},
\qquad 0 \le \Gamma_{ij} \le 1.
\label{eq:Gamma}
\end{equation}
The limiting cases have a direct physical interpretation: $\Gamma_{ij}=1$ yield maximal local coherence, $\Gamma_{ij}=0$ instead refers to complete decoherence. Equation \eqref{eq:Gamma} is directly related to the previously introduced determinant
\begin{equation}
\Gamma_{ij}^2 = 1- \frac{D_{ij}}
{\rho_{ii}\rho_{jj}},
\end{equation}
showing that both quantities encode the same physical information in different normalizations.

\subsubsection*{Three-level principal determinants}

The pairwise determinant naturally extends to larger subsets of levels through the principal minors of the reduced density matrix. Consider a triplet of levels $S=\{i,j,k\}$, and denote by
\begin{equation}
\rho_{ijk}
=
\begin{pmatrix}
\rho_{ii} & \rho_{ij} & \rho_{ik}\\
\rho_{ji} & \rho_{jj} & \rho_{jk}\\
\rho_{ki} & \rho_{kj} & \rho_{kk}
\end{pmatrix}
\end{equation}
the corresponding $3\times3$ principal submatrix. Exploiting Sylvester's criterion\cite{horn2012matrix} for every principal
submatrix of a positive semidefinite matrix,
its determinant satisfies $D_{ijk} = \det(\rho_{ijk})
\ge 0$. Expanding the determinant yields
\begin{equation}
D_{ijk} =\,
\rho_{ii}\rho_{jj}\rho_{kk}
-\rho_{ii}|\rho_{jk}|^2
-\rho_{jj}|\rho_{ik}|^2
-\rho_{kk}|\rho_{ij}|^2 + 2\,\mathrm{Re}
\left(
\rho_{ij}\rho_{jk}\rho_{ki}
\right).
\label{eq:tridet_expanded}
\end{equation}
The first term depends only on the level populations, whereas the second line
contains the genuinely coherent contribution involving the cyclic product of three off-diagonal matrix elements. Unlike the pairwise determinant, the
triplet determinant is therefore sensitive to higher-order coherence among
three levels simultaneously.

Using the Gram representation of the reduced density matrix, $\rho_{mn}=\langle\phi_n|\phi_m\rangle$, the triplet determinant becomes the Gram determinant of the conditional environment states, $D_{ijk} =
\det
\left(
\langle\phi_\beta|\phi_\alpha\rangle
\right)_{\alpha,\beta\in\{i,j,k\}}$. Equivalently,
\begin{equation}
D_{ijk}
=
\left\|
\phi_i
\wedge
\phi_j
\wedge
\phi_k
\right\|^2,
\end{equation}
where $\wedge$ denotes the exterior (wedge) product\cite{bourbaki1989algebra}. The determinant therefore equals the squared volume of the parallelepiped generated by the three conditional states. Consequently,
\begin{itemize}
\item $D_{ijk}=0$ if and only if the three conditional states are linearly
dependent;
\item $D_{ijk}>0$ whenever they span a genuine three-dimensional subspace.
\end{itemize}
The determinant therefore quantifies the amount of independent information that
the environment acquires in order to distinguish simultaneously among the three selected levels.

Since the determinant scales with the total population of the selected levels,
it is often convenient to introduce the normalized triplet determinant
\begin{equation}
\Delta_{ijk} = \frac{D_{ijk}}
{\rho_{ii}\rho_{jj}\rho_{kk}},
\label{eq:normalized_tridet}
\end{equation}
which is well defined whenever
$\rho_{ii}\rho_{jj}\rho_{kk}>0$. Using the normalized conditional states $\ket{\hat{\phi}_i} =
\frac{\ket{\phi_i}}
{\sqrt{\rho_{ii}}}$, Eq.~\eqref{eq:normalized_tridet} can be rewritten as $\Delta_{ijk} =
\det
\left(
\langle
\hat{\phi}_\beta
|
\hat{\phi}_\alpha
\rangle
\right)_{\alpha,\beta\in\{i,j,k\}}$. The normalized determinant satisfies $0\le
\Delta_{ijk} \le 1$, where
\begin{itemize}
\item $\Delta_{ijk}=0$ corresponds to linearly dependent conditional states,
\item $\Delta_{ijk}=1$ corresponds to mutually orthogonal conditional states, which are perfectly distinguishable by the environment.
\end{itemize}

Unlike the raw determinant, the normalized quantity removes the trivial
dependence on the level populations and isolates the intrinsic geometry of the
conditional environment states. It therefore provides a measure of genuine three-level distinguishability, playing for triplets the same role
that the normalized coherence factor $\Gamma_{ij}$ plays for pairs of levels.

\subsubsection*{Conditional pair purity}

A quantitative characterization is obtained by extracting from each
$2\times2$ submatrix associated with a subsystem $S$ the total populations $w = \rho_{ii} + \rho_{jj}$. After normalization, $\widetilde{\rho}_{ij}=\frac{\rho_{ij}}
{w}$, the purity becomes
\begin{equation}
P_S = \mathrm{Tr}(\widetilde{\rho}_S^2) = \frac{\rho_{ii}^2 + \rho_{jj}^2 + 2|\rho_{ij}|^2}
{(\rho_{ii}+\rho_{jj})^2}.
\end{equation}
For a non-empty two-dimensional subspace, $\frac{1}{2} \le P_S \le 1$. The upper bound corresponds to a pure local superposition, while lower values indicate increasing mixedness of the selected pair of levels.

\subsubsection*{Higher-order subspace purities}
The previous construction naturally generalizes to arbitrary subsets of levels.
Let $S=\{i_1,\ldots,i_k\}$, and denote by $\rho_S$ the corresponding principal submatrix of the reduced density matrix. Its total population is $w_S = \mathrm{Tr}(\rho_S)$. Whenever $w_S>0$, the normalized subspace density matrix is $\widetilde{\rho}_S = \frac{\rho_S}{w_S}$, whose purity is
\begin{equation}
P_S = \mathrm{Tr}
\left(
\widetilde{\rho}_S^{2}
\right).
\label{eq}
\end{equation}
These quantities characterize the degree of coherence retained within progressively larger subsets of the Hilbert space.

\bibliography{sample}

%\textcolor{blue}{\noindent LaTeX formats citations and references automatically using the bibliography records in your .bib file, which you can edit via the project menu. Use the cite command for an inline citation, e.g.  \cite{Hao:gidmaps:2014}.}

%For data citations of datasets uploaded to e.g. \emph{figshare}, please use the \verb|howpublished| option in the bib entry to specify the platform and the link, as in the \verb|Hao:gidmaps:2014| example in the sample bibliography file.

\section*{Funding Declaration}

Authors were supported by the Italian funding within the “Budget MUR - Dipartimenti di Eccellenza 2023 - 2027” (Law 232, 11 December 2016) - Quantum Sensing and Modelling for One-Health (QuaSiModO), CUP:H97G23000100001. This paper was funded by the Italian Ministry of University and Research (MUR) within the framework of the National Programme “Research, Innovation and Competitiveness for the Green and Digital Transition” (PN RIC) 2021–2027, Call D.D. No. 307 – Project ECHO TWIN. The project is co-funded by the European Union through the European Regional Development Fund (ERDF). Azione 1.1.2$\_$CUP: B99H26000290007; Azione 1.1.3b$\_$CUP: B92F26000440005; Azione 1.4.3$\_$CUP: B89J26003490005. Funding to Università degli Studi di Bari: Azione 1.1.2$\_$CUP: B99H26000520007; Azione 1.1.3b$\_$CUP: B92F2600053000; Azione 1.4.3$\_$CUP: B99J26001800005.

\section*{Acknowledgements (not compulsory)}

%\textcolor{blue}{Acknowledgements should be brief, and should not include thanks to anonymous referees and editors, or effusive comments. Grant or contribution numbers may be acknowledged.}

Some icons in Figure \ref{fig:gargano} by \url{https://it.vecteezy.com}.

\section*{Author contributions statement}

%\textcolor{blue}{Must include all authors, identified by initials, for example:
%A.A. conceived the experiment(s),  A.A. and B.A. conducted the experiment(s), C.A. and D.A. analysed the results.  All authors reviewed the manuscript.} 

Conceptualization: DP, AC, GRS; Software: DP; Data Curation: AC, GRS; Methodology: DP, AC, GRS; Formal analysis and investigation: DP; Visualization: DP; Funding acquisition: RB, AM, NA; Resources: DP, GRS, EP, AM, NA; Supervision: DP, AM, NA; Writing - original draft preparation: DP, AC, GRS; Writing - review and editing: DP, AC, GRS, LB, DD, ME, AF, FG, NK, RL, EP, ST, RB, AM, NA.

\section*{Additional information}

%\textcolor{blue}{To include, in this order: \textbf{Accession codes} (where applicable); \textbf{Competing interests} (mandatory statement).}

%The corresponding author is responsible for submitting a \href{http://www.nature.com/srep/policies/index.html#competing}{competing interests statement} on behalf of all authors of the paper. This statement must be included in the submitted article file.

The authors declare no competing interests. The embedding data are available through the Google Earth Engine Code Editor at \url{code.earthengine.google.com}. Any additional dataset is available from the corresponding author upon reasonable request.

\end{document}